\documentclass[aps,twocolumn,showpacs]{revtex4}% Physical Review B
\usepackage{graphicx}% Include figure files

\begin{document}
\title{Thermal correction to resistivity in dilute 2D systems}

\author{M. V. Cheremisin}
\affiliation{A.F.Ioffe Physical-Technical Institute,
St.Petersburg, Russia}
\date{\today}

\begin{abstract}
We calculate the resistivity of 2D electron (hole) gas, taking
into account the degeneracy and the thermal correction due to the
combined Peltier and Seebeck effects. The resistivity is found to
be universal function of temperature, expressed in units of
$\frac{h}{e^{2}} (k_{F}l)^{-1}$. The giant parallel
magnetiresistivity found to result from the spin and, if exists,
valley splitting of the energy spectrum. Our analysis of
compressibility and thermopower points to thermodynamic nature of
metal-insulator transition in 2D systems.
\end{abstract}


\maketitle

Recently, a great deal of interest has been focussed on the
anomalous transport behavior of a wide variety of low-density 2D
electron\cite{Kravchenko,Popovic} and
hole\cite{Coleridge,Simmons,Hanein} systems. It has been found
that, below some critical density, the cooling causes an increase
in resistivity, whereas in the opposite high density case the
resistivity decreases. Another unusual property of dilute 2D
systems is their enormous response to parallel magnetic field. At
low temperatures the magnetic field found to suppress the metallic
behavior of 2D electron(hole) gas and result in strong increasing
of resistivity upon enhancement of spin polarization
degree\cite{Okamoto,Vitkalov}. At high temperatures the parallel
magnetoresistivity starts to be unaffected by magnetic field when
the temperature exceeds a value being of the order of Zeeman
energy. A strong perpendicular magnetic field, if applied
simultaneously with the parallel one, results in suppression of
parallel magnetoresistivity\cite{Simonian}. Although numerous
theories have been put forward to account for these effects, the
origin of the above behavior is still the subject of a heated
debate.

The ohmic measurements known to be carried out at low current $I
\rightarrow 0$ in order to prevent heating. Usually, only the
Joule heat is considered to be important. In contrast to the Joule
heat, the Peltier and Thomson effects are linear in current. As
shown in Refs.\cite{Kirby,Cheremisin,Cheremisin+}, the Peltier
effect influences ohmic measurements and results in a correction
to a measured resistance. When current is flowing, one of the
sample contacts is heated, and the other cooled, because of the
Peltier effect. The contact temperatures are different. The
voltage drop across the circuit includes the Seebeck
thermoelectromotive force, which is linear in current. Finally,
there exists a thermal correction $\Delta \rho $, to the ohmic
resistivity, $\rho $, of the sample. For degenerate electrons,
$\Delta \rho /\rho \sim (kT/\mu )^{2}$, where $\mu $ is the Fermi
energy. Hence, the correction may be comparable with the ohmic
resistance of a sample when $kT\sim \mu $.

In the present paper, we report on a study of low-T transport in
2D electron(hole) gas, taking into account both the electron
degeneracy and the Peltier-effect-induced correction to
resistivity.\cite{Kirby,Cheremisin}. The parallel
magnetoresistivity found to originate from the spin and,if exists,
valley splitting of 2D energy spectrum.

\begin{figure}
\vspace*{0.5cm}
\includegraphics[scale=0.38]{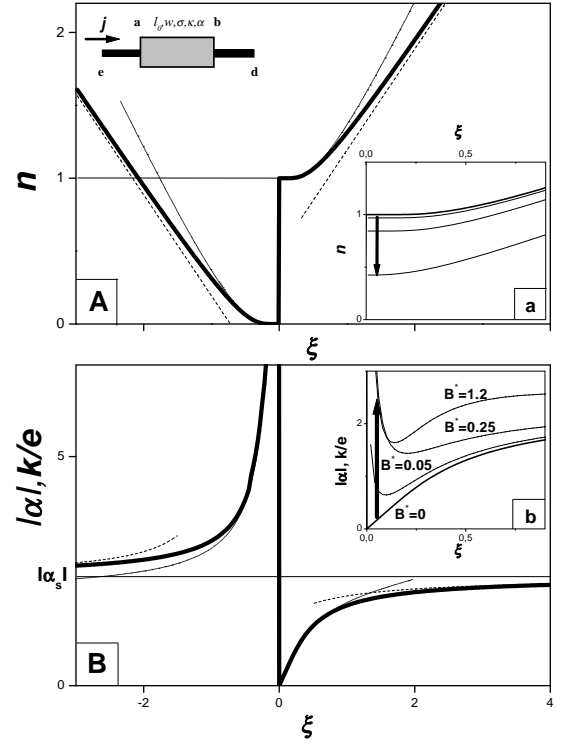}
\caption[]{\label{f.1}
Zero-field 2DEG density(A) and
thermopower(B) given by Eq.(\ref{concentration}) and
Eq.(\ref{alfa}) respectively vs dimensionless temperature $\xi$.
Asymptotes: $\left| \xi \right| \ll 1$ - dotted line, $\xi \gg 1$
- dashed line. Low-temperature dependence of 2DEG density(a) and
thermopower(b) given by Eq.(\ref{N_S_B}) at fixed spin
polarization factor $B^{*}=0;0.05;0.25;1.2$. The arrows depict the
magnetic field enhancement(diminution) of the
thermopower(density). Inset(top panel): experimental setup.}
\vspace*{-0.5cm}
\end{figure}

Let us consider, for clarity, the (100) MOSFET 2DEG system. The
electrons are assumed to occupy the first quantum-well subband
with isotropic energy spectrum $\varepsilon (k)=\frac{\hbar
^{2}{\bf k}^{2}}{2m}$, where $m$ is the effective mass. The sample
is connected (see Fig.\ref{f.1}A, inset) by means of two identical
leads to the current source. Both contacts are ohmic. The voltage
is measured between the open ends( "e" and "d" ) kept at the
temperature of the external thermal reservoir. The sample is
placed in a sample chamber with mean temperature $T_{0}$.

According to our basic assumption, the contacts( "a" and "b" ) may
have different respective temperatures $T_{a}$ and $T_{b}$.
Including the temperature gradient term, the current density ${\bf
j}$ and the energy flux density ${\bf q}$ yield
\begin{equation}
\mathbf {j} = \sigma (\mathbf {E-}\alpha \mathbf{\nabla}T),\qquad
\mathbf {q} = (\alpha T-\zeta /e)\mathbf {j}- \kappa \nabla T,
\label{transport}
\end{equation}
where $\mathbf {E}=\nabla \zeta /e$ is the electric field, and
$\zeta =\mu -e\varphi $ the electrochemical potential. Then,
$\sigma =Ne^{2}\tau /m$ is the conductivity, $N$ the 2DEG density,
$\kappa $ the thermal conductivity, and $\alpha $ the thermopower.

In general, one can unambiguously solve Eq.(\ref{transport}), and,
then find a difference of contact temperatures $\Delta
T=T_{a}-T_{b}$ for an arbitrary circuit cooling. Since the
electron-phonon coupling is weak below $\sim 1$K, the heat
conduction from 2DEG to mixing chamber could predominately occur
through the contacts of the sample and the leads connected to
them. However, the experimental observations\cite{Mittal}
demonstrate that 2DEG alone is the dominant thermal resistance in
this problem. Actually, the cooling is provided by electron
thermal conductivity found to follow Wiedemann-Franz law $\kappa
=LT\sigma$, where $L=\frac{\pi ^{2}k^{2}}{3e^{2}}$ is the Lorentz
number. Accordingly, we further consider adiabatic cooling, with
the 2DEG thermally insulated from the environment. Then, we will
omit the Joule heating for actual $I \rightarrow 0$ case. We
emphasize that under the above conditions, the sample is not
heated. Indeed, at small currents, $T_{a}\approx T_{b}\approx
T_{0}$. Hence, the amount of the Peltier heat, $Q_{a}=I\Delta
\alpha T_{0}$, evolved at contact "a" and that absorbed at contact
"b" are equal. Here, $\Delta \alpha$ is the difference of 2DEG and
metal conductor thermopowers. If it is recalled that the energy
flux is continuous at each contact, the difference of the contact
temperatures yields \cite{Kirby},\cite{Cheremisin} $\Delta
T=\frac {\Delta \alpha l_{0}}{L \sigma w}I$, where $l_{0}$ and $w$
are respectively the sample length and width. Using
Eq.(\ref{transport}), the voltage drop between ends "e" and "d" is
given by $U=RI+\Delta \alpha \Delta T$, where $R$ is the ohmic
resistance of the circuit. The second term denotes conventional
Seebeck thermoelectromotive force. Since $\Delta T \sim I$, we
finally obtain the total 2DEG resistivity as follows
\begin{equation}
\rho ^{tot}=\rho (1+\alpha ^{2}/L) , \label{resistivity}
\end{equation}
where $\rho =1/\sigma $ is the 2DEG ohmic resistivity. In
Eq.(\ref{resistivity}) we ignore conductors resistances and take
into account that for the actual case of the metal leads $\Delta
\alpha \simeq -\alpha $. Eq.(\ref{resistivity}) can also then be
applied for 2D hole gas.

\begin{figure}
\vspace*{0.5cm}
\includegraphics[scale=0.38]{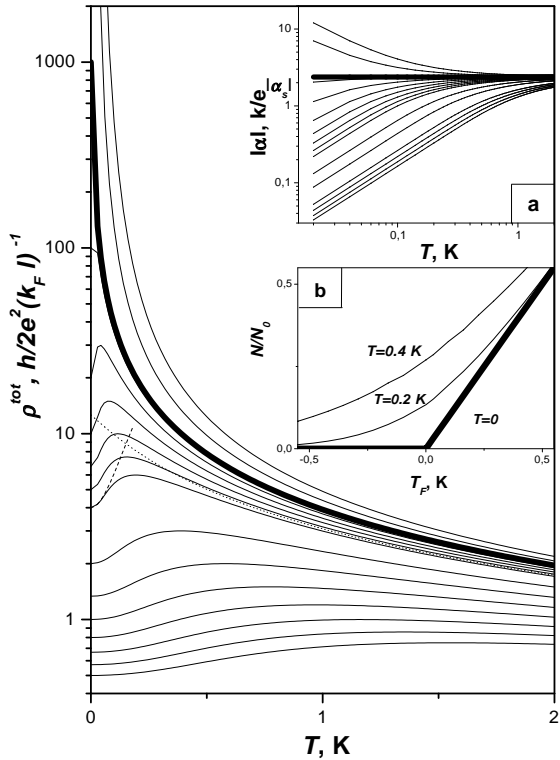}
\caption[]{\label{f.2}
Zero-field T-dependence of resistivity and
thermopower(inset (a)), given by
Eqs.(\ref{resistivity}-\ref{alfa}) for $T_{F}[K]$=2-0.25(step
0.25), 0.2-0.05( step 0.05), 0.01, 0(bold line),-0.1,-0.2.
Asymptotes: $\xi \ll 1$ - dashed line, $ \xi
>1$ - dotted line for fixed $T_{F}=0.25$K. Inset(b):
2DEG density vs Fermi energy at fixed temperature $T=0$ (piecewise
bold line); $0.2; 0.4$K.}
\vspace*{-0.5cm}
\end{figure}

Using Gibbs statistics and the above energy spectrum, the 2DEG
density $N=- {\partial \Omega \overwithdelims()
\partial \mu }_{T}$ yields
\begin{equation}
N=N_{0}\xi F _{0}(1/\xi ), \label{concentration}
\end{equation}
where $\Omega =-kT\sum \ln (1+\exp (\frac{\mu -\epsilon
_{{}}}{kT}))$ is the 2DEG thermodynamic potential, and $\xi
=kT/\mu $ the dimensionless temperature. Then, $N_{0}=D\mu$ is the
density of strongly degenerate 2DEG, $D=\frac{2m}{\pi \hbar ^{2}}$
the 2DEG density of states(factor 2 accounts valley degeneracy)
and $F _{n}(z)$ the Fermi integral. In Fig.\ref{f.1}A, we plot the
temperature dependence of the dimensionless concentration
$n=N/N_{0}$ given by Eq.(\ref{concentration}). In the classical
Maxwell-Boltzman limit ($\xi <0,\left| \xi \right| \ll 1$), the 2D
electron density is thermally activated, and, therefore, $n=\left|
\xi \right| \exp (-1/\left| \xi \right| )$. In the case of
strongly degenerate electrons ($\xi \ll 1$), we obtain $n=1+\xi
\exp (-1/\xi )$. Then, at elevated temperatures $\xi \geq 1$, the
dependence of the 2DEG density $n=1/2+\xi \ln 2$ becomes linear in
temperature. It is to be noted that at finite temperature the 2DEG
density always exceeds( see Fig.\ref{f.2}, inset b) the
zero-temperature value, i.e $N>N_{0}$. As a consequence, the
density, deduced from the period of Shubnikov-de-Haas quantum
oscillations, $N_{SdH}=N_{0}$, is less than that, $N_{Hall}=N$,
obtained from low-field Hall resistivity data( see
Ref.\cite{Cheremisin++}) in consistent with
experiments.\cite{Pudalov}
We emphasize that the 2D electron density is a monotonic function
of temperature (see Fig.\ref{f.1}A). Therefore, one might expect
that the ohmic resistivity $\rho (T)\sim 1/N$ decreases with
increasing temperature at constant carrier mobility. We now
demonstrate that the total resistivity specified by
Eq.(\ref{resistivity}) can, nevertheless, increase in a certain
temperature range owing to thermal correction.

Following the conventional Boltzman equation formalism, the
explicit formulae for the 2DEG thermopower (for the 3D case, see
Pisarenko, 1940) can be written as
\begin{equation}
\alpha =-\frac{k}{e}\left[ \frac{2F _{1}(1/\xi )}{F _{0}(1/\xi
)}-\frac{1}{\xi }\right]. \label{alfa}
\end{equation}
Here, we assume, for simplicity, that the electron scattering is
characterized by energy-independent momentum relaxation time. In
the classical Maxwell-Boltzman limit ($\xi <0,\left| \xi \right|
\ll 1$) the thermopower is given by the conventional formulae
$\alpha =-\frac{k}{e}(2-1/\xi )$. Then, for strongly degenerated
2DEG ($\xi \ll 1$), we obtain the temperature dependence of the
thermopower (Fig.\ref{f.1}B) as $\alpha =-\frac{k}{e}[\pi ^{2}\xi
/3-(1+3\xi )\exp (-1/\xi )]$. At elevated temperatures ($\xi >1$)
the thermopower first grows with temperature, and then approaches
the universal value $\alpha _{s}=-\frac{k}{e}\frac{2F _{1}(0)}{F
_{0}(0)}=-\frac{k}{e}\frac{\pi ^{2}}{6\ln 2}$. Our support of the
above behavior is confirmed by low-T thermopower measurements
data\cite{Fletcher}, found to diverge at certain value $\sim
0.6k/e$ being of the order of $\alpha _{s}$( see bold line in
Fig.\ref{f.2}, inset a).

It's worth noting that the thermopower can be of the order of
$k/e$. Accordingly, the thermal correction to resistivity may
comparable with the ohmic resistivity of 2DEG. In Fig.\ref{f.2},
we plot the temperature dependence of the 2DEG resistivity given
by Eq.(\ref{resistivity}) at different Fermi temperatures $
T_{F}=\mu /k$. At fixed temperature, the resistivity increases
with decreasing of 2DEG degeneracy.  At certain Fermi energy
($T_{F}=0.25$K in Fig.\ref{f.2}) the T-dependence of the
resistivity exhibits metallic behavior at $T<T_{F}$, and then
becomes insulating ( i.e. $\frac{d\rho }{dT} <0$ ) at $T>T_{F}$.
Within the low-temperature metallic region $\xi \ll 1$, the 2DEG
resistivity can be approximated (see dashed line in Fig.\ref{f.2}
) with $\rho ^{tot}=\rho _{0}(1+\pi ^{2}\xi ^{2}/3)$, where $\rho
_{0} =\frac{h}{2e^{2}}(k_{F}l)^{-1}$ is the resistivity at
$T\rightarrow 0$, $k_{F}=\sqrt{2m\mu }/\hbar $ the Fermi vector,
and $l=$ $\hbar k_{F}\tau /m$ the mean free path. Then, for the
high-temperature ($\xi >1$)
insulating region we obtain the asymptote $\rho ^{tot}=\rho _{0}\frac{%
1+\alpha _{s}^{2}/L}{\xi \ln 2+1/2}$, depicted in Fig.\ref{f.2} by
dotted line. This results are confirmed by recent
experiments\cite{Hamilton,Lewalle} showing that for temperatures
well below the Fermi temperature the metallic region data obey a
scaling, where the disordered parameter $k_{F}l$ and dimensionless
temperature $T/T_{F}$ appear explicitly. These experimental
observations\cite{Hamilton} therefore rule out interactions, the
shape of the potential well, spin-orbit effects and quantum
interference effects\cite{Brunthaler} as possible origins of the
metallic behavior mechanism. It is to be noted that the
conventional theory\cite{Zala} used to explain 2D metallic
behavior\cite{Pudalov4} is, however, failed to account both $T
\rightarrow 0$ and $T \geq T _{F}$ cases.
It is of particular interest the 2DEG compressibility,
$K=\frac{dN}{d\mu}=-\frac{d^2 \Omega}{d \mu^2}$, known to be a
fundamental quantity generally more amenable to theoretical and
experimental analysis.\cite{Smith,Eisenstein,Dultz} For
noninteracting 2DEG system Eq.(\ref{concentration}) yields
$K(\xi)=DF'_{0}(1/\xi)$, where $F'_{n}(z)=\frac{dF_{n}(z)}{dz}$ is
the derivative of the Fermi function.
\begin{figure}
\vspace*{0.5cm}
\includegraphics[scale=0.38]{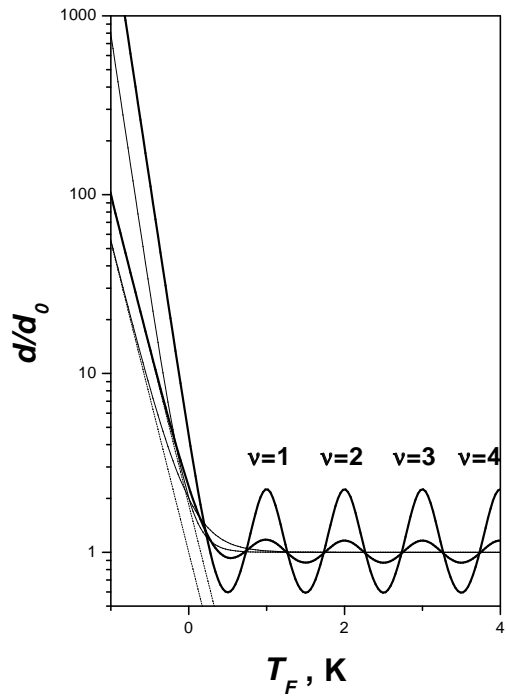}
\caption[]{\label{f.3} Dimensionless inverse compressibility vs
Fermi temperature at zero magnetic field(thin lines) and
$\hbar\omega_{c}=1$K(bold lines) at fixed temperatures
$T=0.15,0.25$K. Dashed lines depict asymptotes at $\xi<0, \left|
\xi \right| \ll 1$} \vspace*{-0.5cm}
\end{figure}
At fixed temperature Fig.\ref{f.3} represents the dependence of
the inverse compressibility parameter, $d(\mu)=\epsilon/Ke^2$, vs
Fermi temperature. For strongly degenerated electrons($\xi \ll 1$)
one obtains a constant value $d_{0}=\epsilon/De^{2}$ in consistent
with experimental findings.\cite{Eisenstein} Then, upon decreasing
of 2DEG density( i.e. $\mu \rightarrow 0$) the experimental
data\cite{Smith,Eisenstein,Dultz} exhibit diminish and,
furthermore, the negative inverse compressibility compared to
$d_{0}$. Usually, this behavior is explained\cite{Eisenstein} in
terms of conventional Hartree-Fock exchange omitted in our simple
approach. However, for extremely low 2D density the inverse
compressibility data always exhibits an abrupt upturn which cannot
be explained\cite{Dultz} within Hartree-Fock scenario. We argue
that the above feature has the natural explanation within our
model(see dashed line in Fig.\ref{f.3}) since
$d=d_{0}exp(-1/\left| \xi \right|)$ at $\xi<0, \left| \xi \right|
\ll 1$ and, hence, exhibits T-activated behavior.

In general case of 2DEG placed in strong perpendicular magnetic
field $B_\perp$ the compressibility yields
\begin{widetext}
\begin{equation}
K=\frac{D}{4 \xi \nu}\sum \limits_{n}
\frac{1}{\cosh\left(\frac{\varepsilon_{n} - \mu}{2kT}\right)^{2}}
\simeq D \left [F'_{0}(1/\xi) +4 \pi^{2} \xi \nu \sum\limits_{k}
\frac{(-1)^{k}k \cos(2 \pi \nu k)}{ \sinh(2\pi^{2}k \nu \xi)}
\right ]. \label{compressibility}
\end{equation}
\end{widetext}
Here, $\Omega=-kT \Gamma \sum \limits_{n} \ln \left(1+\exp
\left(\frac{\mu -\varepsilon_{n}}{kT}\right)\right)$ is the
thermodynamic potential modified with respect to 2DEG
spin-unresolved Landau level(LL) energy spectrum
$\varepsilon_{n}=\hbar\omega_{c}(n+1/2)$, where $n=0,1..$ is the
LL number, $\omega_{c}= \frac{eB_\perp}{mc}$ the cyclotron
frequency, and $\Gamma$ the zero-width LL density of states. Then,
$\nu=\frac{\mu}{\hbar \omega_{c}}$ is the filing factor. According
to Eq.(\ref{compressibility}), at fixed magnetic field the
dependence $d(\mu)$ can be viewed( see Fig.\ref{f.3}) as a
superposition of zero-field dependence and LLs related
oscillations. Then, at $\xi<0, \left| \xi \right| \ll 1$ in
presence of strong magnetic field $\hbar\omega_{c}>>kT$(
i.e.$\xi\nu<<1$) the inverse compressibility exhibits T-activated
behavior as $d=d_{0}\xi\nu \exp \left( \frac{1}{2\xi
\nu}-\frac{1}{\left| \xi \right|} \right)$ similar to zero-field
case( see Fig.\ref{f.3}). These findings are consistent with
experimental observations.\cite{Eisenstein,Dultz}

\begin{figure}
\vspace*{0.5cm}
\includegraphics[scale=0.38]{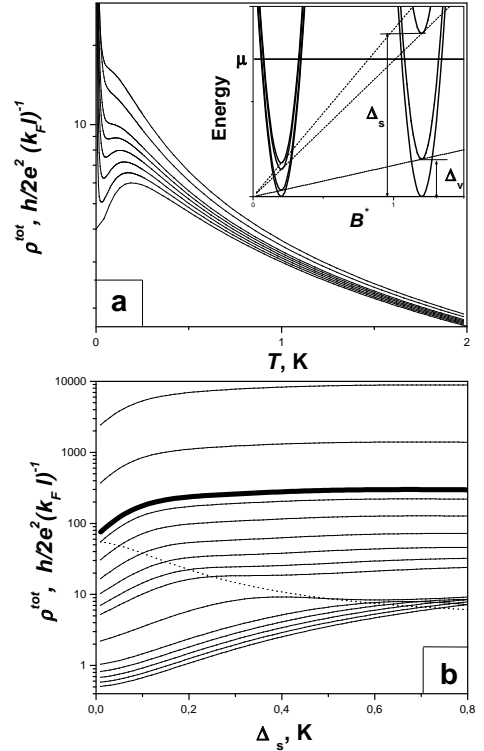}
\caption[]{\label{f.4} (a)T-dependence of magnetoresistivity at
fixed $T_{F}=0.25K$ for different degree of spin polarization(
from top to bottom) $B^{*}=0.2; 0.15; 0.1-0.02$(step $0.02$); $0$.
Inset: energy diagram for valley and(or) spin subbands.
(b)Magnetoresistivity vs Zeeman splitting at fixed temperature
$T=0.07$K for Fermi energies denoted in Fig.\ref{f.2}. Dotted line
represents the condition $B^{*}=1$ for spin-polarized 2DEG.}
\vspace*{-0.5cm}
\end{figure}

We now address the question of parallel field magnetoresistivity.
The energy spectrum modified with respect to valley
\cite{Pudalov3}
$\Delta_{v}[K]=\Delta_{v}^{0}+0.6B_{\parallel}[T]$ and spin
$\Delta_{s}[K]=g^{*}\mu_{B}B_{\parallel}=2.6B_{\parallel}[T]$
splitting is represented in Fig.\ref{f.4}a, inset. For simplicity,
we neglect further the density dependence of $g^{*}$-factor. Then,
addressing the unresolved problem to extract zero-field valley
splitting within low-field Shubnikov de-Haas measurements, we
assume that $\Delta_{v}^{0}=0$. Actually, these assumptions do not
substantially affect our basic results. Note that the lowest
spin-up states are always valley-split since the we have a ratio
$\beta=\Delta_{v}/\Delta_{s}=0.23<1$.

With the help of Gibbs approach, the explicit formulae for 2DEG
density and thermopower yields
\begin{eqnarray}
N= \frac{N_{0} \xi}{4} \sum\limits_{i} F _{0} \left(
\frac{1-E_{i}}{\xi} \right),\qquad
\eqnum{5} \label{N_S_B} \\
\alpha=-\frac{k}{e} \left[ \frac {2\sum\limits_{i} \left[ F _{1}
\left( \frac{1-E_{i}}{\xi} \right)+\frac {E_{i}}{\xi} F _{0}
\left( \frac {1-E_{i}}{\xi} \right) \right]} {\sum\limits_{i} F
_{0} \left( \frac{1-E_{i}}{\xi} \right)} - \frac{1}{\xi }\right],
\nonumber
\end{eqnarray}
where $E_{i}=0,B^{*},\beta B^{*}, (1+\beta) B^{*}$ is the energy
deficit between the bottom of spin and(or) valley subbands and
that of the ground state( see Fig.\ref{f.4}a, inset),
$B^{*}=\Delta_{s}/\mu$ the spin polarization factor. The system
becomes completely spin-polarized when $B^{*}=1$(at $T=0$).
Let us discuss qualitatively the 2DEG magnetoresitivity. At fixed
temperature $\xi \ll 1$ the total number of electrons decreases(
see Fig.\ref{f.1}a) upon increasing of the parallel magnetic field
as $n=1-(1+\beta)B^{*}/2$ when $B^{*}< 1$. Simultaneously, the
thermopower exhibits a strong enhancement( see Fig.\ref{f.1}b)
when $\xi< B^{*}$. We argue, that both the enhancement of
thermopower and diminish of 2DEG density is responsible for the
observed giant magnetoresistivity in 2DEG(Fig.\ref{f.4}b). In the
opposite spin-polarized case $B^{*}\ge 1$, the 2D electrons occupy
the lowest valley-split subbands(Fig.\ref{f.4}a, inset), therefore
$n=1/2-\beta B^{*}/4 \sim 1/2$. Then, for spin-polarized system
the thermopower saturates( see Fig.\ref{f.1}b), and, hence results
in subsequent saturation of the parallel magnetoresistivity(see
Fig.\ref{f.4}b). It worthwhile to mention that the strong parallel
magnetoresistivity observed in p-GaAs/AlGaAs system \cite{Yoon}
can be explained within our model as well. However, the high-field
behavior of resistivity remains unclear.

At fixed magnetic field the T-dependence of magnetoresistivity
given by Eqs.(\ref{resistivity}),(\ref{N_S_B}) is similar to that
represented in Fig.\ref{f.2} for B=0 case. Then, for fixed Fermi
temperature( see Fig.\ref{f.4}a) the above dependence becomes
unaffected by applied magnetic field above certain temperature
being of the order of Zeeman energy. This finding is consistent
with experimental observations.\cite{Simonian} It is worth to be
noted that the conventional theory\cite{Zala} agrees with
magnetoresistivity data\cite{Pudalov4} only qualitatively.

Let us analyze in more detail the cooling conditions, which are
known to influence the thermal correction to
resistivity.\cite{Kirby,Cheremisin} It will be recalled that in
the case of adiabatic cooling the electron temperature differs
from the bath temperature $T_{0}$. We now consider the opposite
situation of electron cooling due to, for example, finite strength
of electron-phonon coupling. Following Ref.\cite{Prus}, below
$\sim $0.6K in Si MOSFET's the electron-to-phonon thermal exchange
is given by $a(T^{3}-T_{0}^{3})$, where $a=$2.2 $\times
$10$^{-8}$W/K$^{3}$cm$ ^{2}$. When $T-T_{0}\ll T_{0}$, the thermal
correction to the resistivity $ \Delta \rho $ is
suppressed\cite{Cheremisin} by the factor $ \gamma =\frac{\tanh
\lambda }{\lambda }$, where $\lambda =l_{0}T_{0}\sqrt{ 3a/4\kappa
}$ is a dimensionless parameter. Actually, $\lambda $ is the ratio
of outgoing and internal heat fluxes associated with phonon
related thermal leakage and electron heat diffusion, respectively.
When $\lambda \ll 1$, the local cooling due to phonons can be
neglected and the adiabatic approach is well justified. In the
opposite case of intensive cooling ($\lambda \gg 1$), the
difference of the contact temperatures $ \Delta T$ becomes
smaller, and, therefore, the thermal correction to resistivity
$\Delta \rho $ vanishes. For $l_{0}=1$mm, $T_{0}=50$mK, $\sigma
=2e^{2}/h=8\times 10^{-5}$Ohm$^{-1}$( typical critical region
conductance) we obtain $\lambda =1.2$, and, therefore, $\gamma
=0.8$. It worthwhile to notice that Peltier correction to
resistivity becomes greater at ultra-low temperatures for short
samples, since $\lambda \sim l_{0}T_{0}$.

We emphasize that both dc and ac ohmic measurements lead to a
thermal correction. The correction is, however, strongly damped at
high frequencies because of the thermal inertial effects. As
demonstrated in Ref.\cite{Kirby}, the above quasi-static
approach is valid below some critical frequency $f_{cr}=\chi
/l_{0}^{2}$. For example, for degenerate electrons the thermal
diffusion coefficient $\chi $ is of the order of the diffusion
coefficient $ D=\frac{\sigma }{e^{2}}\left( \frac{dN_{0}}{d\mu
}\right) ^{-1}$. Assuming $ \sigma =e^{2}/h$, $l_{0}=1$mm, for
GaAs-based structure we obtain $\chi \sim \hbar /m$, hence
$f_{cr}=1.5$kHz. We suggest that the spectral dependence of the 2D
resistivity can be used to estimate the thermal correction.

In conclusion, low-temperature ohmic measurements of a 2D electron
(hole) gas involve a thermal correction caused by the Peltier
effect. The resistivity of 2DEG with thermal correction included
is found to be universal function of temperature, expressed in
units $h/e^{2}(k_{F}l)^{-1}$. This universal behavior correlates
with that found in experiments. Strong increasing, and, then
subsequent saturation in parallel magnetiresistivity in Si-MOSFET
found to results from both the spin and valley splitting of 2DEG
energy spectrum compared with $B=0$ case. We suggest ac
measurements as a powerful tool illuminating the importance of
thermal correction to resistivity. Our analysis of compressibility
and thermopower points to thermodynamic nature of metal-insulator
transition in 2D systems.

This work was supported by RFBR, INTAS(YSF 2001/1-0132), and the
LSF(HPRI-CT-2001-00114, Weizmann Institute)

\bibliography{cond-mat}

\end{document}